\documentclass[preprint,eqsecnum,preprintnumbers,nofootinbib,b61080yrevtex,prd,aps,showpacs,showkeys,groupedaddress,floatfix]{revtex4}
\usepackage{bm}
\usepackage{graphics}
\usepackage{graphicx}
\usepackage{epsfig}
\usepackage{amssymb}
\usepackage{amsmath}

\begin{document}
\title{MESON WAVE FUNCTIONS FROM HOLOGRAPHIC QCD AND THE ROLE OF INFRARED  RENORMALONS IN PROTON-PROTON COLLISIONS}
\author{A.~I.~Ahmadov$^{1,2}$~\footnote{E-mail: ahmadovazar@yahoo.com}}
\author{C.~Aydin$^{1}$~\footnote{E-mail: coskun@ktu.edu.tr}}
\author{F.~Keskin$^{1}$~\footnote{E-mail: feridunkeskin@ktu.edu.tr}}

\affiliation {$^{1}$ Department of Physics, Karadeniz Technical
University, 61080, Trabzon, Turkey\\
$^{2}$ Department of Theoretical Physics, Baku State University, Z.
Khalilov st. 23, AZ-1148, Baku, Azerbaijan}
\begin{abstract}
We calculate the contribution of the higher twist Feynman diagrams
to the large-$p_T$  inclusive pion production cross section in
proton-proton collisions in the case of the running and frozen
coupling approaches within holographic QCD. The structure of
infrared renormalon singularities of the higher twist subprocess
cross section and it's resummed expression are found. We compare the
resummed higher twist cross sections with the ones obtained in the
framework of the frozen coupling approximation and leading twist
cross section. We discuss the phenomenological consequences of
possible higher-twist contributions to the pion production in
proton-proton collisions within holographic QCD.
\end{abstract}
\pacs{12.38.-t, 13.60.Le, 14.40.Aq, 13.87.Fh, } \keywords{higher
twist, holographic QCD, infrared renormalons,} \maketitle

\section{\bf Introduction}
One of the most interesting new developments in hadron physics is
the application of the AdS/CFT correspondence~\cite{Maldacena} to
nonperturbative QCD
problems~\cite{Polchinski,Janik,Erlich,Karch,Brodsky1,Teramond1,Teramond2}.
It is well know that AdS/CFT  gives an important insight into the
viscosity and other global properties of the hadronic system
formed in heavy ion collisions~\cite{Kovtun}. The essential ansatz
for the application of AdS/CFT to hadron physics is the indication
that the QCD coupling $\alpha_s(Q^2)$ becomes large. Therefore
conformal symmetry can be applied to, for example solutions of the
QCD Dyson Schwinger equations  and phenomenological studies of QCD
couplings based on physical observables such as $\tau$ decay and
the Bjorken sum rule show that the QCD $\beta$ function vanishes
and $\alpha_s(Q^2)$ become constant at small virtuality; i.e.,
effective charges develop an infrared fixed point. Fully
exploiting the gauge/gravity correspondence to produce a model for
real strong interaction physics- a method called " holographic
QCD" or "AdS/QCD"-may be attempted either through a top-dawn
approach starting with a particular string theory and choosing a
background that naturally produces QCD-like properties, or a
bottom-up approach starting with real QCD properties and using
them to obtain constraints on viable dual gravity theories.

The first attempts have been made for constructing phenomenological
holographic models of QCD~\cite{Erlich}. Surprisingly simple models
consisting of gauge theory in an anti-de-Sitter space interval have
turned out to provide a remarkably good description of the meson
sector of QCD. Therefore, it will be interesting that the
calculation of the higher twist effects within holographic QCD in
proton-proton collisions in the running coupling approach.

The large-order behavior of a perturbative expansion in gauge
theories is inevitably dominated by the factorial growth of
renormalon diagrams~\cite{Hooft,Mueller,Zakharov,Beneke}. In the
case of quantum chromodynamics (QCD), the coefficients of
perturbative expansions in the QCD coupling $\alpha_{s}$ can
increase dramatically even at low orders. This fact, together with
the apparent freedom in the choice of renormalization scheme and
renormalization scales, limits the predictive power of perturbative
calculations, even in applications involving large momentum
transfer, where $\alpha_{s}$ is effectively small.

In this work we apply the running coupling approach~\cite{Agaev}
in order to compute the effects of the infrared renormalons on the
pion production in proton-proton collisions within holographic
QCD. This approach was also employed
previously~\cite{Ahmadov3,Ahmadov4,Ahmadov5,Ahmadov6} to calculate
the inclusive meson production in proton-proton and photon-photon
collisions.

For the calculations of the higher-twist cross sections on the
dependence of wave functions of pion, we used the holographic QCD
prediction $\Phi_{hol}(x)$~\cite{Brodsky2,Brodsky3,Vega} and pion
asymptotic wave functions $\Phi_{asy}(x)$~\cite{Lepage1} from the
perturbative QCD evolution. Theoretically obtained predicted
results within holographic QCD were compared with results of the
perturbative QCD which obtained by the running coupling and frozen
coupling constants approaches.

The frozen coupling constant approach in
Refs.~\cite{Bagger,Baier,Ahmadov1,Ahmadov2} was used for calculation
of integrals, such as
\begin{equation}
I\sim \int\frac{\alpha_{s}({Q}^2)\Phi(x,{Q}^2)}{1-x}dx.
\end{equation}
It should be noted  that, in pQCD calculations, the argument of the
QCD coupling constant (or the renormalization and factorization
scale) ${Q}^2$ should be taken equal to the square of the momentum
transfer of a hard gluon in a corresponding Feynman diagram. But
definition of $\alpha_{s}(\hat{Q}^2)$ suffers from infrared
singularities. Therefore, in the soft regions as $x_{1}\rightarrow
0$, and  $x_{2}\rightarrow 0$, integrals (1.1) diverge and for their
calculation some regularization methods are needed for
$\alpha_{s}(Q^2)$ in these regions. Investigation of the infrared
renormalon effects in various inclusive and exclusive processes is
one of the most important and interesting problem in the
perturbative QCD. It is known that infrared renormalons are
responsible for factorial growth of coefficients in perturbative
series for the physical quantities. But, these divergent  series can
be resummed by means of the Borel transformation ~\cite{Hooft} and
the principal value prescription ~\cite{Contopanagos}. Studies of
higher-twist and renormalon effects also opened  new prospects for
evaluation of power suppressed corrections to processes
characteristics.

We organize the  paper as the follows. In Section \ref{ht}, we
provide some formulas for the calculation of the contributions of
the higher twist  and leading twist diagrams. In Section \ref{ir},
we present the formulas and analysis of the higher-twist effects
on the dependence of the pion wave function by the running
coupling constant approach, and in Section \ref{results}, the
numerical results for the cross section and discuss the dependence
of the cross section on the pion wave functions are presented.
Finally, some concluding remark are stated in Section \ref{conc}.

\section{HIGHER TWIST AND LEADING TWIST CONTRIBUTIONS TO INCLUSIVE REACTIONS}
\label{ht} The higher-twist Feynman diagrams, which describe the
subprocess $q_1+\bar{q}_{2} \to \pi^{+}(\pi^{-})+\gamma$ for the
pion production in the proton-proton collision are shown in Fig.1.
The amplitude for this subprocess can be found by means of the
Brodsky-Lepage formula ~\cite{Lepage2}
\begin{equation}
M(\hat s,\hat
t)=\int_{0}^{1}{dx_1}\int_{0}^{1}dx_2\delta(1-x_1-x_2)\Phi_{\pi}(x_1,x_2,Q^2)T_{H}(\hat
s,\hat t;x_1,x_2).
\end{equation}
In Eq.(2.1), $T_H$ is  the sum of the graphs contributing to the
hard-scattering part of the subprocess.

The Mandelstam invariant variables for subprocesses $q_1+\bar{q}_{2}
\to \pi^{+}(\pi^{-})+\gamma$ are defined as
\begin{equation}
\hat s=(p_1+p_2)^2,\quad \hat t=(p_1-p_{\pi})^2,\quad \hat
u=(p_1-p_{\gamma})^2.
\end{equation}
The pion wave functions predicted by
AdS/QCD~\cite{Brodsky2,Brodsky3,Vega} and the PQCD evolution
~\cite{Lepage1} has the form:
$$
\Phi_{asy}^{hol}(x)=\frac{4}{\sqrt{3}\pi}f_{\pi}\sqrt{x(1-x)},
$$
\begin{equation}
\Phi_{VSBGL}^{hol}(x)=\frac{A_1k_1}{2\pi}\sqrt{x(1-x)}exp\left(-\frac{m^2}{2k_{1}^2x(1-x)}\right),\quad
\Phi_{asy}^{p}(x)=\sqrt{3}f_{\pi}x(1-x)
\end{equation}
where $f_{\pi}$ is the pion decay constant.

The cross section for the higher-twist subprocess $q_1\bar{q}_{2}
\to \pi^{+}(\pi^{-})\gamma$ is given by the expression
\begin{equation}
\frac{d\sigma}{d\hat t}(\hat s,\hat t,\hat u)=\frac
{8\pi^2\alpha_{E} C_F}{27}\frac{\left[D(\hat t,\hat
u)\right]^2}{{\hat s}^3}\left[\frac{1}{{\hat u}^2}+\frac{1}{{\hat
t}^2}\right]
\end{equation}
where
\begin{equation}
D(\hat t,\hat u)=e_1\hat
t\int_{0}^{1}dx\left[\frac{\alpha_{s}(Q_1^2)\Phi_{\pi}(x,Q_1^2)}{1-x}\right]+e_2\hat
u\int_{0}^{1}dx\left[\frac{\alpha_{s}(Q_2^2)\Phi_{\pi}(x,Q_2^2)}{1-x}\right].
\end{equation}
In the Eq.(2.5) $Q_{1}^2=(x-1)\hat u \,\,\,\,$and $Q_{2}^2=-x\hat
t$ \,\, represent the momentum squared carried by the hard gluon
in Fig.1, $e_1(e_2)$ is the charge of $q_1(\overline{q}_2)$ and
$C_F=\frac{4}{3}$. The higher-twist contribution to the
large-$p_{T}$ pion production cross section in the process
$pp\to\pi^{+}(\pi^{-})+\gamma+X$ is ~\cite{Owens,Greiner}
\begin{equation}
\Sigma_{M}^{HT}\equiv E\frac{d\sigma}{d^3p}=\int_{0}^{1}\int_{0}^{1}
dx_1 dx_2 G_{{q_{1}}/{h_{1}}}(x_{1})
G_{{q_{2}}/{h_{2}}}(x_{2})\frac{\hat s}{\pi} \frac{d\sigma}{d\hat
t}(q\overline{q}\to \pi\gamma)\delta(\hat s+\hat t+\hat u).
\end{equation}
We denote the higher-twist cross section obtained using the frozen
coupling constant approach by $(\Sigma_{\pi}^{HT})^0$.

Regarding the higher-twist corrections to the pion production
cross section, a comparison of our results with leading-twist
contributions is crucial. We take two leading-twist subprocesses
for the pion production:(1) quark-antiquark annihilation $q\bar{q}
\to g\gamma$, in which the  $g \to \pi^{+}(\pi^{-})$ and (2)
quark-gluon fusion, $qg \to q\gamma $, with subsequent
fragmentation of the final quark into a meson, $q \to
\pi^{+}(\pi^{-})$ ~\cite{Ahmadov3,Ahmadov5}.

\section{THE HIGHER TWIST MECHANISM IN HOLOGRAPHIC QCD AND INFRARED RENORMALONS}\label{ir}

The main problem in our investigation is the calculation of integral
in (2.5) by the running coupling constant approach within
holographic QCD and also discussion of the problem of normalization
of the higher twist process cross section in the context of the same
approach. Therefore, it is worth noting that, the renormalization
scale (argument of $\alpha_s$) according to Fig.1 should be chosen
equal to $Q_{1}^2=(x-1)\hat u$, $Q_{2}^2=-x\hat t$. The integral in
Eq.(2.5) in the framework of the running coupling approach takes the
form
\begin{equation}
I(\mu_{R_{0}}^2)=\int_{0}^{1}\frac{\alpha_{s}(\lambda
\mu_{R_0}^2)\Phi_{M}(x,\mu_{F}^2)dx}{1-x}.
\end{equation}
The $\alpha_{s}(\lambda \mu_{R_0}^2)$ has the infrared singularity
at $x\rightarrow1$, for $\lambda=1-x$  or $x\rightarrow0$, for
$\lambda=x$ and so the integral $(3.1)$ diverges. For the
regularization of the integral, we express the running coupling at
scaling variable $\alpha_{s}(\lambda \mu_{R_0}^2)$ with the aid of
the renormalization group equation in terms of the fixed one
$\alpha_{s}(Q^2)$. The solution of renormalization group equation
for the running coupling $\alpha\equiv\alpha_{s}/\pi$ has the form
~\cite{Contopanagos}
\begin{equation}
\frac{\alpha(\lambda)}{\alpha}=\left[1+\alpha
\frac{\beta_{0}}{4}\ln{\lambda}\right]^{-1}.
\end{equation}
Then, for $\alpha_{s}(\lambda Q^2)$, we get
\begin{equation}
\alpha(\lambda Q^2)=\frac{\alpha_{s}}{1+\ln{\lambda/t}}
\end{equation}
where $t=4\pi/\alpha_{s}(Q^2)\beta_{0}=4/\alpha\beta_{0}$.

Having inserted  Eq.(3.3) into Eq.(2.5) we obtain
$$
D(\hat t,\hat u)=e_{1}\hat t\int_{0}^{1}dx\frac{\alpha_{s}(\lambda
\mu_{R_0}^2)\Phi_{M}(x,Q_{1}^2)}{1-x}+ e_{2}\hat
u\int_{0}^{1}dx\frac{\alpha_{s}(\lambda
\mu_{R_0}^2)\Phi_{M}(x,Q_{2}^2)}{1-x}
$$
\begin{equation}
=e_{1}\hat t\alpha_{s}(-\hat u)t_{1}\int_{0}^{1}dx
\frac{\Phi_{M}(x,Q_{1}^2)}{(1-x)(t_{1}+\ln\lambda)} + e_{2}\hat
u\alpha_{s}(-\hat t)t_{2}\int_{0}^{1}dx
\frac{\Phi_{M}(x,Q_{2}^2)}{(1-x)(t_{2}+\ln\lambda)}
\end{equation}
where $t_1=4\pi/\alpha_{s}(-\hat u)\beta_{0}$ and
$t_2=4\pi/\alpha_{s}(-\hat t)\beta_{0}$.

Although the integral (3.4) is still divergent, it is recast into a
suitable  form for calculation. Making the change of variable as
$z=\ln\lambda$, we obtain
\begin{equation}
D(\hat t,\hat u)=e_{1}\hat t \alpha_{s}(-\hat u) t_1\int_{0}^{1}dx
\frac{\Phi_{M}(x,Q_{1}^2)}{(1-x)(t_1+z)}+ e_{2}\hat u
\alpha_{s}(-\hat t) t_2 \int_{0}^{1} dx
\frac{\Phi_{M}(x,Q_{2}^2)}{(1-x)(t_2+z)}
\end{equation}
In order to calculate (3.5) we will apply the integral
representation of $1/(t+z)$ ~\cite{Zinn-Justin,Erdelyi}.
\begin{equation}
\frac{1}{(t+z)}=\int_{0}^{\infty}e^{-(t+z)u}du,
\end{equation}
gives
\begin{equation}
D(\hat t,\hat u)=e_{1} \hat{t} \alpha_{s}(-\hat u) t_1 \int_{0}^{1}
\int_{0}^{\infty} \frac{\Phi_{\pi}(x,Q_{1}^2)e^{-(t_1+z)u}du
dx}{(1-x)}+ e_{2} \hat{u} \alpha_{s}(-\hat t) t_2 \int_{0}^{1}
\int_{0}^{\infty} \frac{\Phi_{\pi}(x,Q_{2}^2)e^{-(t_2+z)u}du
dx}{(1-x)}
\end{equation}
In the case $\Phi_{asy}^{hol}(x)$ for the $D(\hat t,\hat u)$ it is
written as
\begin{equation}
D(\hat t,\hat u)=\frac{16 f_{\pi} e_{1} \hat t}{\sqrt{3}\beta_{0}}
\int_{0}^{\infty} du
e^{-t_{1}u}B\left(\frac{3}{2},\frac{1}{2}-u\right)+  \frac{16
f_{\pi} e_{2} \hat u}{\sqrt{3}\beta_{0}} \int_{0}^{\infty} du
e^{-t_{2}u}B\left(\frac{3}{2},\frac{1}{2}-u\right)
\end{equation}
and for $\Phi_{asy}^{p}(x)$ wave function
\begin{equation}
D(\hat t,\hat u)=\frac{4\sqrt{3}\pi f_{\pi}e_{1}\hat t}{\beta_{0}}
\int_{0}^{\infty}du e^{-t_{1}u}
\left[\frac{1}{1-u}-\frac{1}{2-u}\right] +\frac{4\sqrt{3}\pi
f_{\pi}e_{2}\hat u}{\beta_{0}} \int_{0}^{\infty}du e^{-t_{2}u}
\left[\frac{1}{1-u}-\frac{1}{2-u}\right].
\end{equation}
where $B(\alpha,\beta)$ is Beta function.  The structure of the
infrared renormalon poles in Eq.(3.8) and Eq.(3.9) strongly depend
on the wave functions of the pion. To remove them from Eq.(3.8) and
Eq.(3.9) we adopt the principal value prescription. We denote the
higher-twist cross section obtained using the running coupling
constant approach by $(\Sigma_{\pi}^{HT})^{res}$.

\section{NUMERICAL RESULTS AND DISCUSSION}\label{results}

In this section, we discuss the higher-twist contributions
calculated in the context of the running and frozen coupling
constant approaches on the dependence of the chosen pion wave
functions in the process $pp \to \pi^{+}(or\,\, \pi^{-})\gamma+X$.
In numerical calculations for the quark distribution function
inside the proton, the MSTW distribution function ~\cite{Martin},
and the gluon and quark fragmentation ~\cite{Albino} functions
into a pion have been used. The results of our numerical
calculations are displayed in Figs.2-14. Firstly, it is very
interesting comparing the higher-twist cross sections obtained
within  holographic QCD with the ones obtained within perturbative
QCD. In Fig.2 and Fig.3 we show the dependence of higher-twist
cross sections $(\Sigma_{\pi^{+}}^{HT})^{0}$,
$(\Sigma_{\pi^{+}}^{HT})^{res}$ calculated in the context of the
frozen and  running coupling constant approaches as a function of
the pion transverse momentum $p_{T}$ for different pion wave
functions at $y=0$. It is seen from Fig.2 and Fig.3 that the
higher-twist cross section is monotonically decreasing with an
increase in the transverse momentum of the pion. In Fig.4-Fig.7,
we show the dependence of the ratios
$(\Sigma_{HT}^{hol})$/$(\Sigma_{HT}^p)$,
$(\Sigma_{\pi}^{HT})^{res}$/$(\Sigma_{\pi^{+}}^{HT})^{0}$,
$(\Sigma_{\pi^{+}}^{HT})^{0}$/$(\Sigma_{\pi^{+}}^{LT})$ and
$(\Sigma_{\pi^{+}}^{HT})^{res}$/$(\Sigma_{\pi^{+}}^{LT})$ as a
function of the pion transverse momentum $p_{T}$ for
$\Phi_{\pi}^{hol}(x)$, $\Phi_{\pi}^{p}(x)$ and
$\Phi_{VSBGL}^{hol}(x)$ pion wave functions. Here
$\Sigma_{\pi^{+}}^{LT}$ is the leading-twist cross section,
respectively. As shown in Fig.4, in the region
$2\,\,GeV/c<p_T<30\,\,GeV/c$ resummed higher-twist cross section
for $\Phi_{\pi}^{hol}(x)$ is suppress by about half orders of
magnitude relative to the resummed higher-twist cross section for
$\Phi_{\pi}^{p}(x)$ also higher-twist cross section for
$\Phi_{\pi}^{p}(x)$ is suppress by about half orders of magnitude
relative to the  higher-twist cross section for
$\Phi_{VSBGL}^{hol}(x)$ . In Fig.5 and Fig.6, we show the
dependence of the ratios
$(\Sigma_{\pi}^{HT})^{res}$/$(\Sigma_{\pi^{+}}^{HT})^{0}$, and
$(\Sigma_{\pi^{+}}^{HT})^{0}$/$(\Sigma_{\pi^{+}}^{LT})$ as a
function of the meson transverse momentum $p_{T}$ for the
$\Phi_{\pi}^{hol}(x)$,  $\Phi_{\pi}^{p}(x)$ and
$\Phi_{VSBGL}^{hol}(x)$ pion wave functions. It is observed from
Fig.5 that, the ratios  for all wave functions decrase with an
increase in the transverse momentum of pion, but in the region
$2\,\,GeV/c<p_T<8\,\,GeV/c$ resummed higher-twist cross section
for $\Phi_{\pi}^{hol}(x)$ is suppress by about 2-3 and 2-4 orders
of magnitude relative to the higher-twist cross sections
calculated in the context of the frozen coupling method for
$\Phi_{\pi}^{p}(x)$ and $\Phi_{VSBGL}^{hol}(x)$, respectively.
Also, resummed higher-twist cross section for $\Phi_{\pi}^{p}(x)$
is suppress by about 1-3 orders of magnitude relative to the
higher-twist cross sections calculated in the context of the
frozen coupling method for $\Phi_{VSBGL}^{hol}(x)$. But, as shown
in Fig.6, the ratios decrease with increasing in the $p_{T}$
transverse momentum of the pion and has a minimum approximately at
the point $p_{T}=13 GeV/c$, then the ratios increase with
increasing in the $p_{T}$ transverse momentum of the pion. In
Fig.7 is similar to Fig.6 with an exception that there in a
minimum approximately at the point $p_{T}=25 GeV/c$. In Fig.8 -
Fig.10, we have depicted higher-twist cross sections
$(\Sigma_{\pi^{+}}^{HT})$, and ratio
$(\Sigma_{HT}^{hol})/(\Sigma_{HT}^{p})$,\, as a function of the
rapidity $y$ of the pion at $\sqrt s=62.4\,\,GeV$ and
$p_T=4.9\,\,GeV/c$. Figures show that higher-twist cross section
and ratios have a different distinctive. The resummed higher-twist
cross section for $\Phi_{\pi}^{hol}(x)$ has a maximum
approximately at the point $y=-1.92$. However in this point
higher-twist cross section for $\Phi_{VSBGL}^{hol}(x)$ has a
minimum. As shown in Fig.9 in the region ($-2.52\leq y\leq -1.92$)
the ratios $(\Sigma_{HT}^{hol})^{0}/(\Sigma_{HT}^{p})^{0}$ and
$(\Sigma_{VSBGL}^{hol})^{0}/(\Sigma_{HT}^{p})^{0}$ increase with
an increase of the $y$ rapidity of the pion and has a maximum
approximately at the point $y=1.22$, but ratio
$(\Sigma_{HT}^{hol})^{0}/(\Sigma_{VSBGL}^{hol})^{0}$ has a minimum
in this point. As is seen from Fig.10 ratios
$R=(\Sigma_{\pi^{+}}^{HT})^{res}/(\Sigma_{\pi^{+}}^{HT})^{0}$,\,
for all wave functions increase with an increase of the $y$
rapidity of the pion and has a maximum approximately at the point
$y=-1.92$. In the region  $-2.52<y<-1.92$ resummed higher-twist
cross section is suppress by about half order of magnitude
relative to the higher-twist cross section calculated in the
framework of the frozen coupling approach. Besides that, the ratio
decreases with an increase in the $y$ rapidity of the pion. As is
seen from Fig.10, the ratio $R$  depends on the choice of the pion
wave function. Analysis of our calculations shows that
$(\Sigma_{\pi^{+}}^{HT})^{0}$, $(\Sigma_{\pi^{+}}^{HT})^{res}$
higher-twist cross sections and ratio sensitive to pion wave
functions predicted holographic and perturbative QCD.

We have also carried out comparative calculations in the
center-of-mass energy $\sqrt s=200\,\,GeV$ and obtained results
are displayed in Fig.11-Fig.14.  Analysis of our calculations at
the center-of-mass energies $\sqrt s=62.4\,\,GeV$ and $\sqrt
s=200\,\,GeV$, show that with increasing in the beam energy
contributions of higher-twist effects  to the cross section
decrease by about one-two order. As is seen from Fig.5, Fig.7,
Fig.10 and Fig.13 that infrared renormalon effects enhance the
perturbative predictions for the pion production cross section in
the proton-proton collisions about 2-3 order. This feature of
infrared renormalons  may help the explain theoretical
interpretations with future experimental  data for the pion
production cross section in the proton-proton collisions. In our
calculations, the higher-twist cross section of the process the
dependence of the transverse momentum of pion appears in the range
of $(10^{-8}\div10^{-26})mb/GeV^2$. Therefore, higher-twist cross
section obtained in our work should be observable at RHIC.

\section{CONCLUSIONS}\label{conc}

Proton-proton collisions are known  are known to be the most
elementary interactions and form the very basis of our knowledge
about the nature of high energy collisions in general. Physicists,
by and large, hold the view quite firmly that the perturbative
quantum-chromodynamics  provides a general framework for the
studies on high energy particle-particle collisions. Obviously,
the unprecedented high energies attained at Large Hadron Collider
offer new window and opportunities to test the proposed QCD
dynamics with its pros and cons. However, we should remember that
LHC opens a new kinematical regime ah high energy, where several
questions related to the description of the high-energy regime of
the QCD. Consequently, studies of proton-proton interactions at
the RHIC  and LHC could provide valuable  information on the QCD
dynamics at high energies. In this work the single meson inclusive
production via higher twist mechanism within holographic QCD are
calculated . For calculation of the cross section, the running
coupling constant approach is applied and infrared renormalon
poles in the cross section expression are revealed. Infrared
renormalon induced divergences is regularized by means of the
principal value prescripton and the resummed expression (the Borel
sum) for the higher-twist cross section is found. It is observed
that, the resummed higher-twist cross section differs considerably
from that found using the frozen coupling approximation, in some
region. The following results can be concluded from the
experiments; the higher-twist contributions to single meson
production cross section in the proton-proton collisions have
important phenomenological consequences, the higher-twist pion
production cross section in the proton-proton collisions depends
on the form of the pion model wave functions and may be used for
their study. Also the contributions of renormalons effects within
holograpich QCD in this process are essential and may help to
analyse experimental results. Further investigations are needed in
order to clarify the role of higher-twist effects  in QCD.
Especially, the forthcoming RHIC and LHC measurements will provide
further tests of the dynamics of large-$p_T$ hadron production
beyond the leading twist.

\section*{Acknowledgments}
One of author A.I.Ahmadov is grateful to all members of the
Department of Physics of Karadeniz Technical University for
appreciates hospitality extended to him in Trabzon. Financial
support by TUBITAK under grant number 2221(Turkey) is also
gratefully acknowledged.


\newpage
\begin{figure}[!hbt]
\vskip 1.2cm \epsfxsize 16cm \centerline{\epsfbox{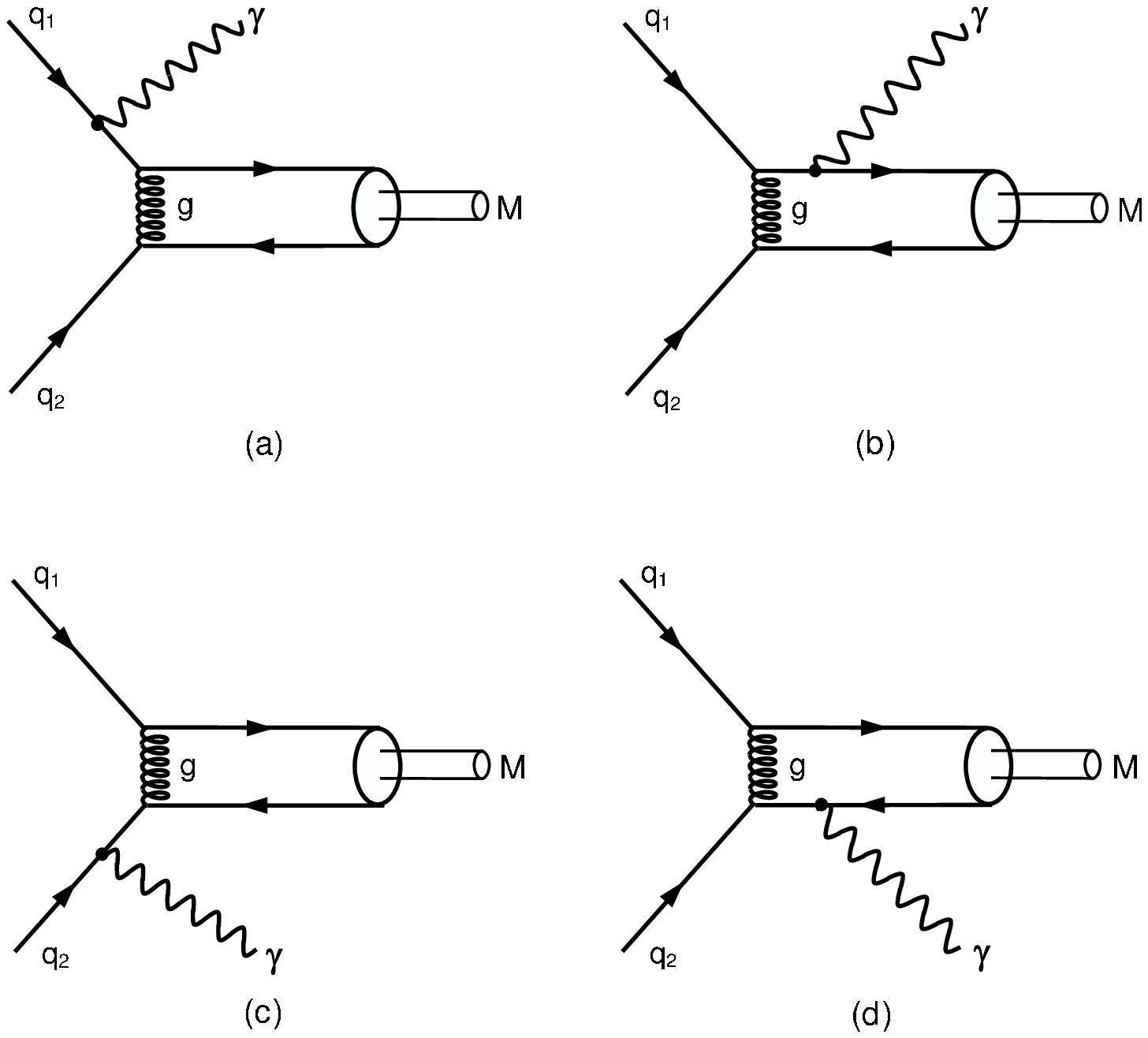}} \vskip
-5cm \caption{Feynman diagrams for the  higher twist subprocess,
$q_1 q_2 \to \pi^{+}(or\,\,\pi^{-})\gamma.$} \label{Fig1}
\end{figure}


\begin{figure}[!hbt]
\vskip -1.2cm\epsfxsize 11.8cm \centerline{\epsfbox{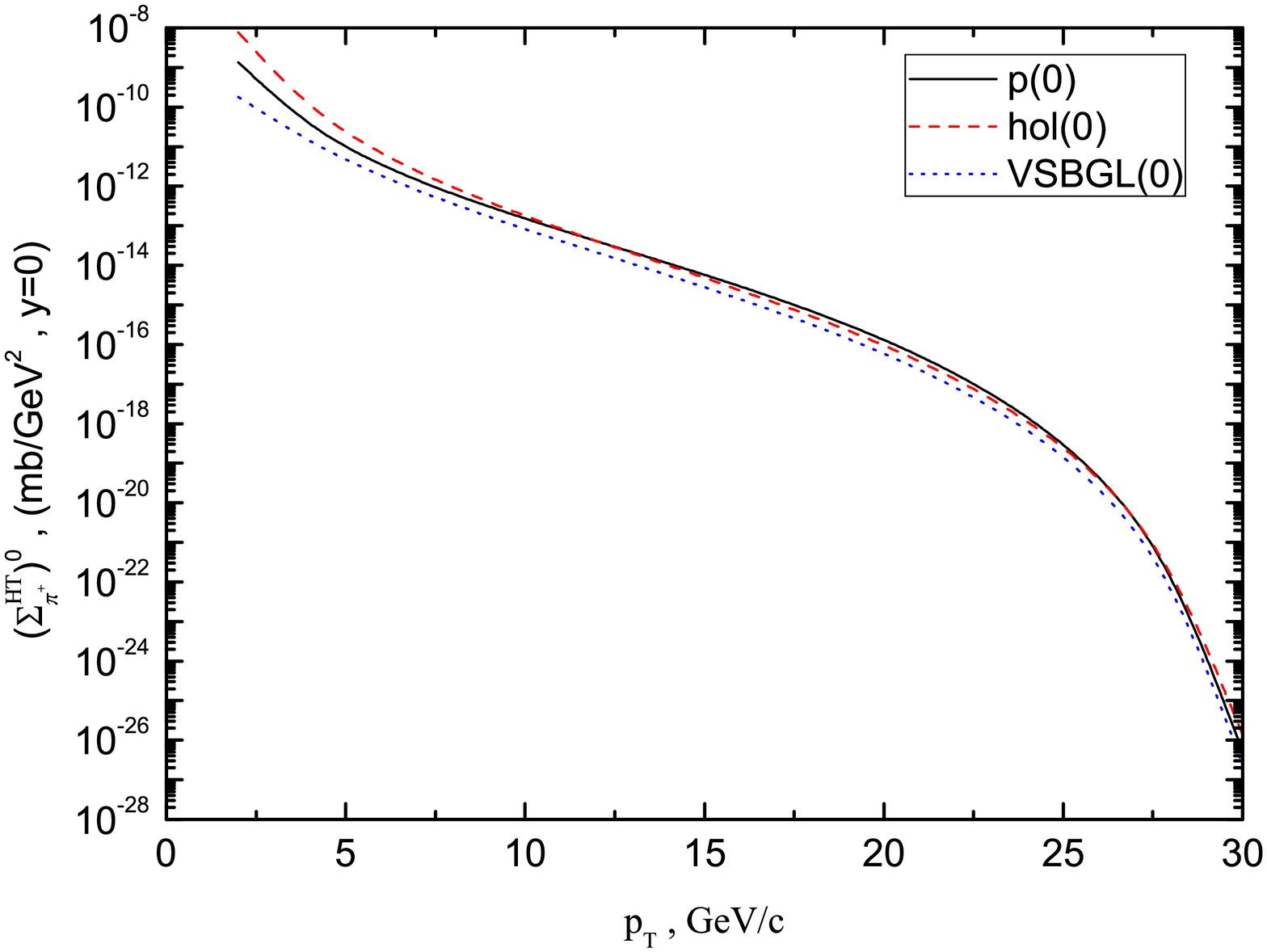}}
\vskip-0.2cm \caption{Higher-twist $\pi^{+}$ production cross
section $(\Sigma_{\pi^{+}}^{HT})^{0}$ as a function of the $p_{T}$
transverse momentum of the pion at the c.m. energy $\sqrt s=62.4\,\,
GeV$.} \label{Fig2}
\end{figure}

\begin{figure}[!hbt]
\vskip 1.2cm \epsfxsize 11.8cm \centerline{\epsfbox{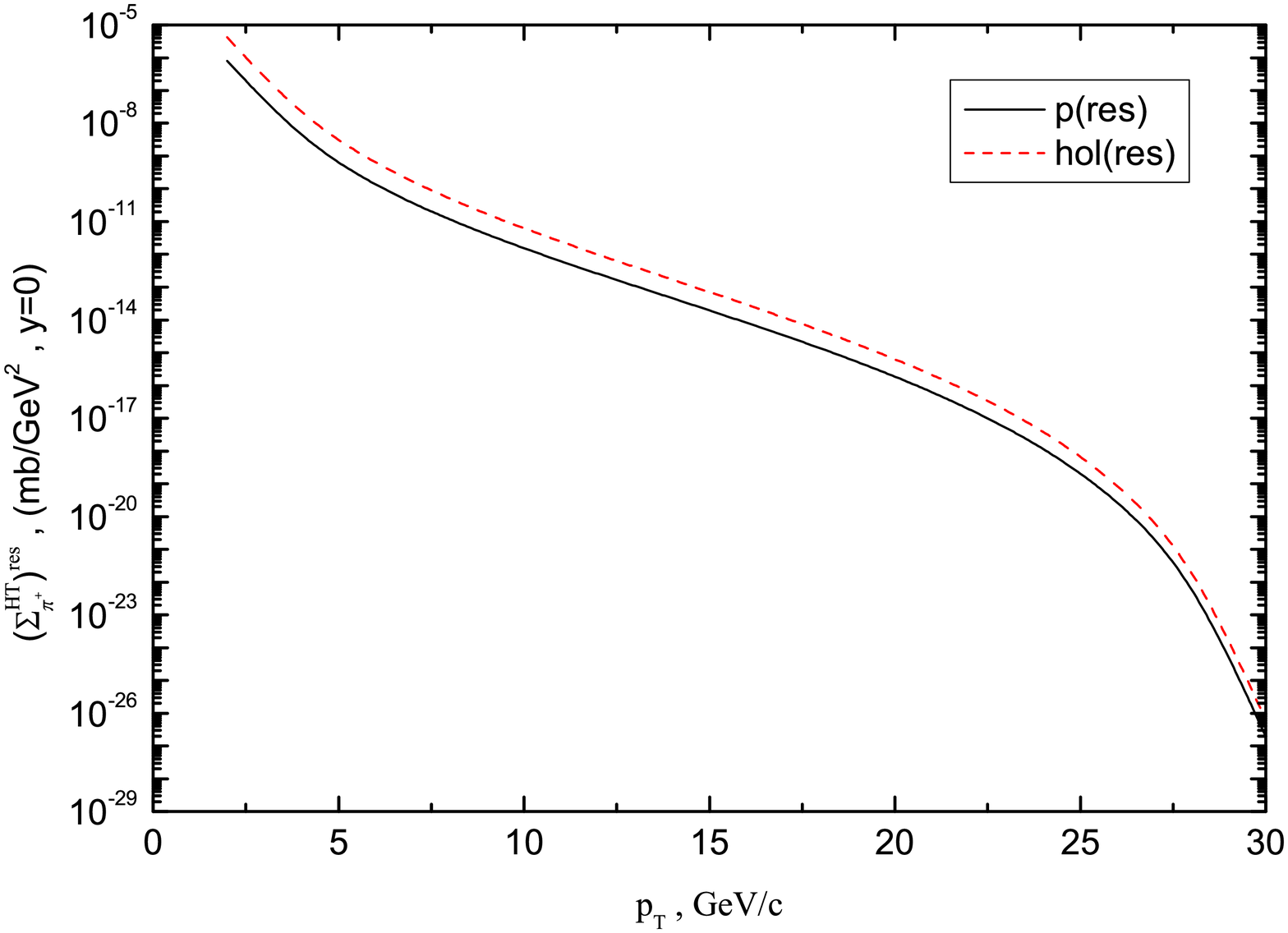}}
\vskip-0.2cm \caption{Higher-twist $\pi^{+}$ production cross
section $(\Sigma_{\pi^{+}}^{HT})^{res}$ as a function of the $p_{T}$
transverse momentum of the pion at the c.m.energy $\sqrt s=62.4\,\,
GeV$.} \label{Fig3}
\end{figure}

\begin{figure}[!hbt]
\vskip -1.2cm\epsfxsize 11.8cm \centerline{\epsfbox{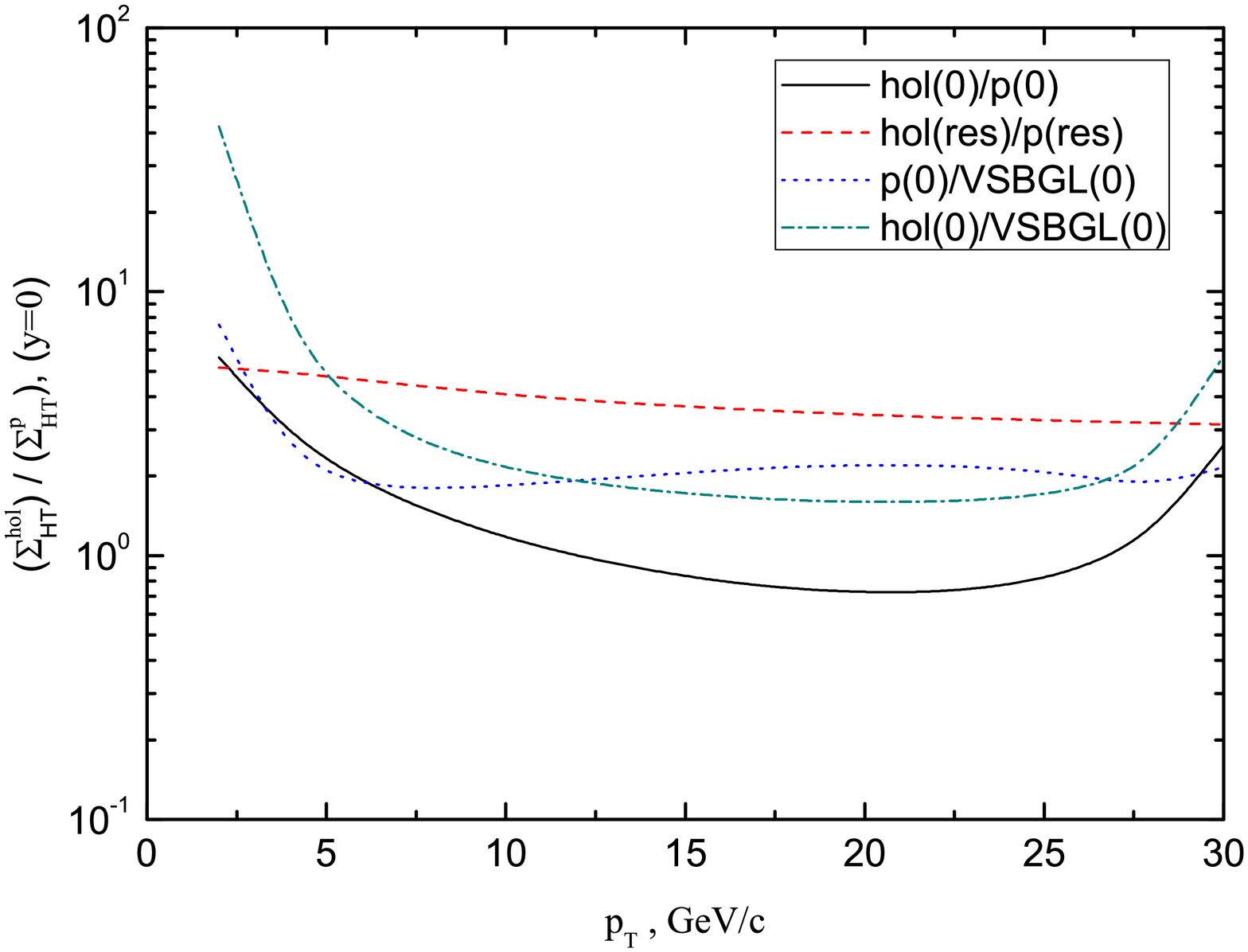}}
\vskip-0.2cm \caption{Ratio $(\Sigma_{HT}^{hol})/(\Sigma_{HT}^p)$,
where higher-twist contribution are calculated for the pion rapidity
$y=0$ at the c.m.energy $\sqrt s=62.4\,\, GeV$ as a function of the
pion transverse momentum, $p_{T}$.} \label{Fig4}
\end{figure}

\begin{figure}[!hbt]
\vskip 1.2cm\epsfxsize 11.8cm \centerline{\epsfbox{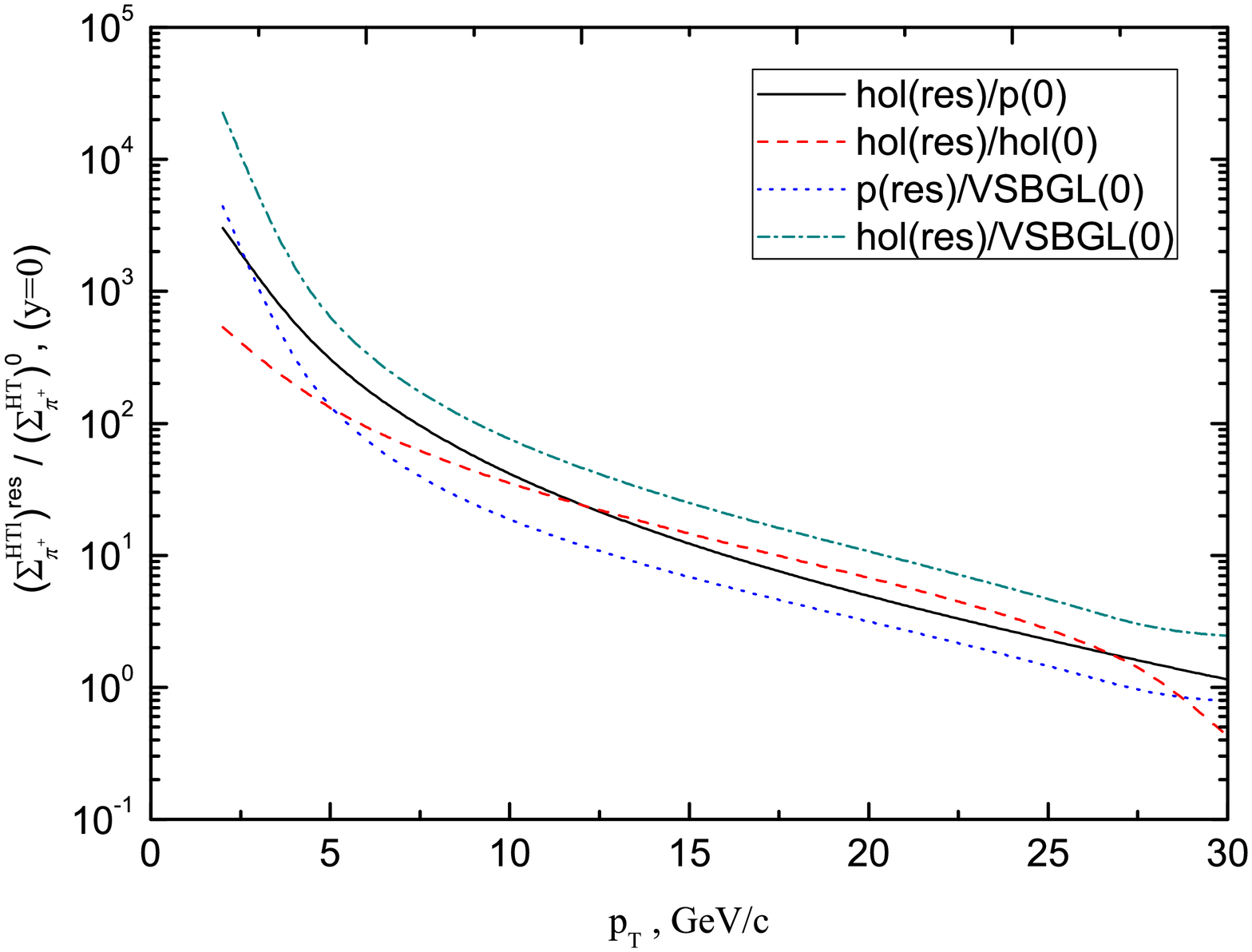}}
\vskip-0.2cm \caption{Ratio
$(\Sigma_{\pi^{+}}^{HT})^{res}/(\Sigma_{\pi^{+}}^{HT})^0$, as a
function of the $p_{T}$ transverse momentum of the pion at the c.m.
energy $\sqrt s=62.4\,\, GeV$.} \label{Fig5}
\end{figure}

\begin{figure}[!hbt]
\vskip -1.2cm\epsfxsize 11.8cm \centerline{\epsfbox{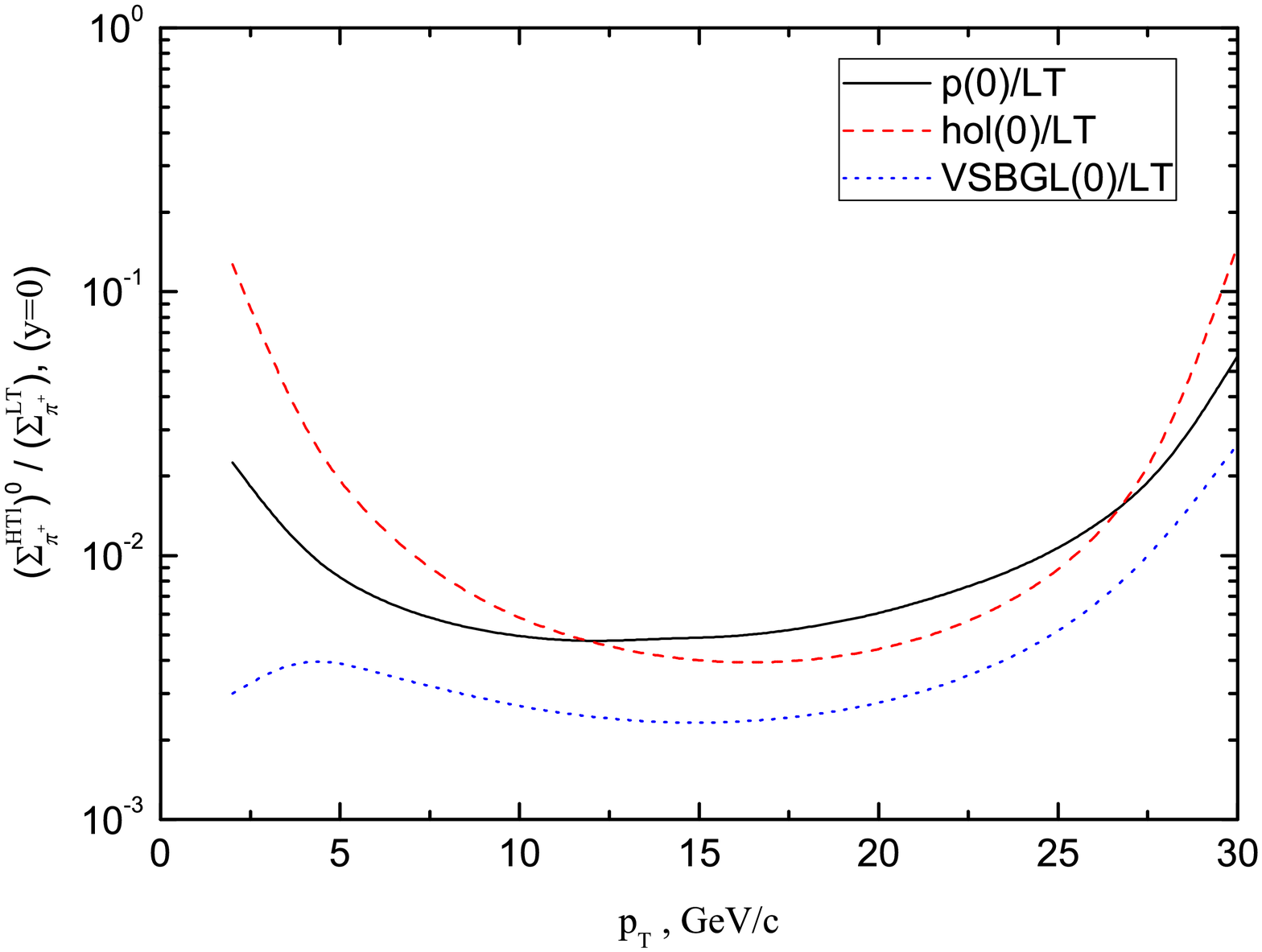}}
\vskip-0.2cm \caption{Ratio
$(\Sigma_{\pi^{+}}^{HT})^0/(\Sigma_{\pi^{+}}^{LT})$, where
higher-twist contribution are calculated for the pion rapidity $y=0$
at the c.m.energy $\sqrt s=62.4\,\, GeV$ as a function of the pion
transverse momentum, $p_{T}$.} \label{Fig6}
\end{figure}

\begin{figure}[!hbt]
\vskip 1.2cm\epsfxsize 11.8cm \centerline{\epsfbox{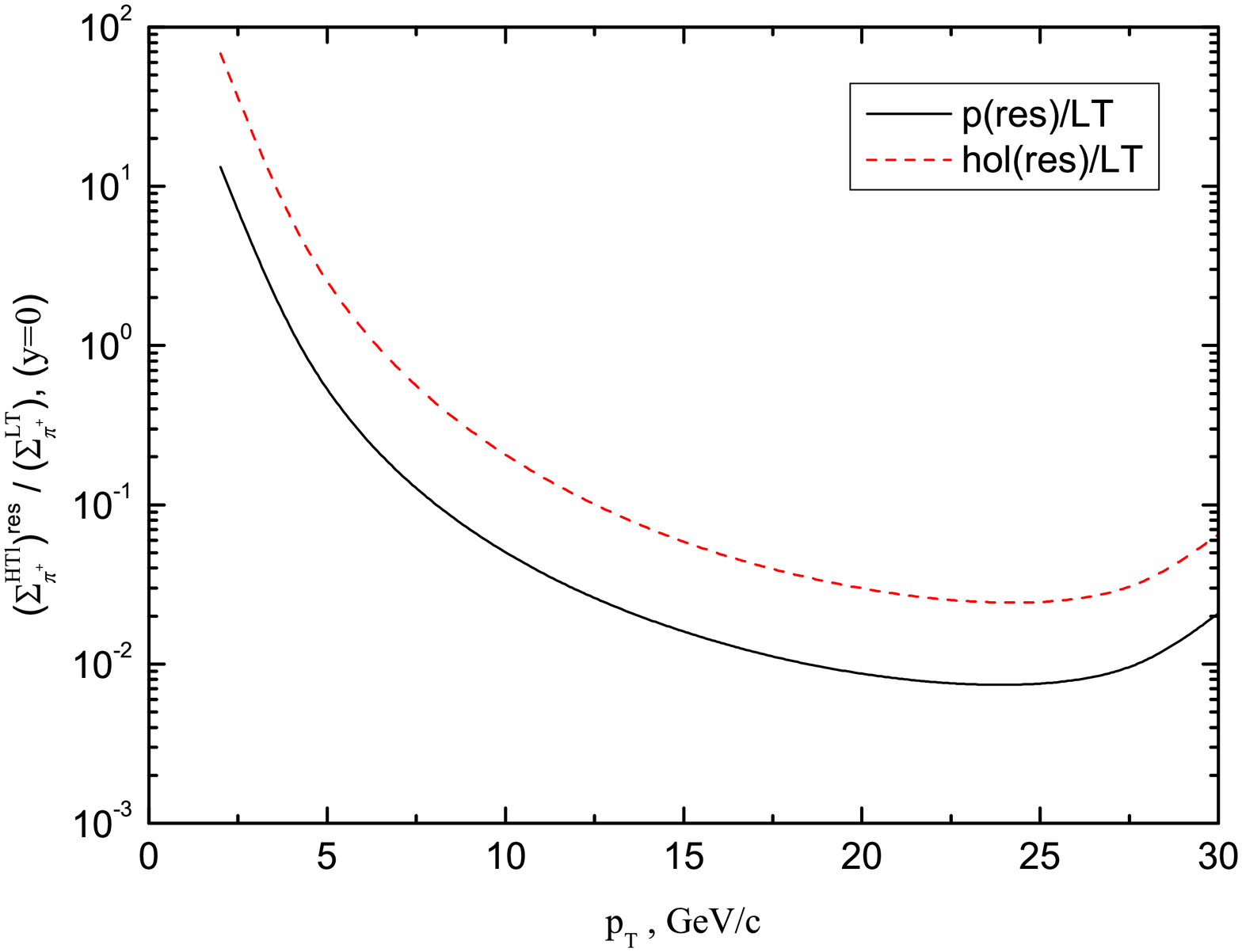}}
\vskip-0.2cm \caption{Ratio
$(\Sigma_{\pi^{+}}^{HT})^{res}/(\Sigma_{\pi^{+}}^{LT})^0$, as a
function of the $p_{T}$ transverse momentum of the pion at the c.m.
energy $\sqrt s=62.4\,\, GeV$.} \label{Fig7}
\end{figure}

\begin{figure}[!hbt]
\vskip-1.2cm \epsfxsize 11.8cm \centerline{\epsfbox{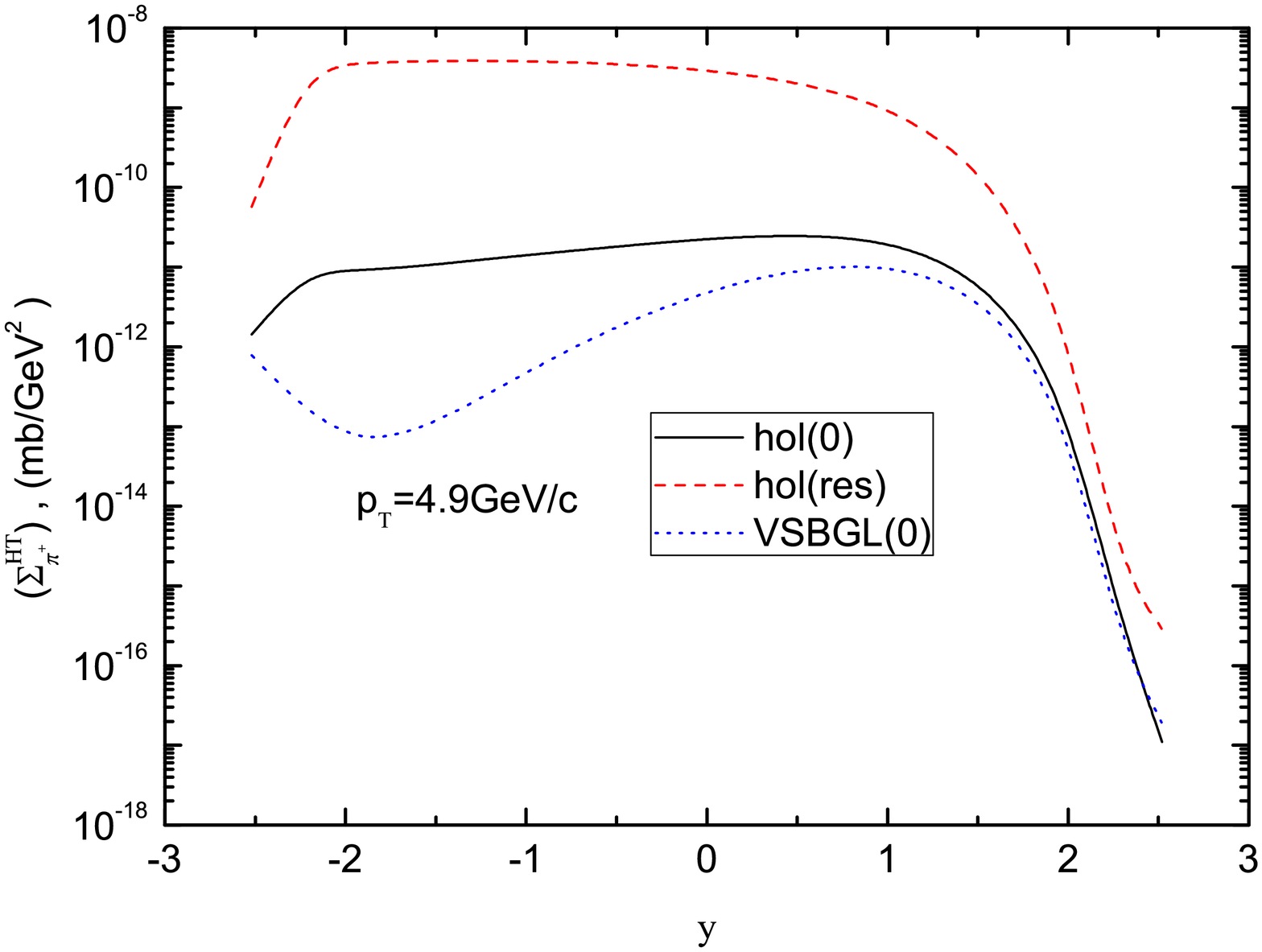}}
\vskip-0.2cm \caption{Higher-twist $\pi^{+}$ production cross
section  $(\Sigma_{\pi^{+}}^{HT})$ , as a function of the $y$
rapidity of the pion at the  transverse momentum of the pion
$p_T=4.9\,\, GeV/c$, at the c.m. energy $\sqrt s=62.4\,\, GeV$.}
\label{Fig8}
\end{figure}

\begin{figure}[!hbt]
\vskip 0.8cm \epsfxsize 11.8cm \centerline{\epsfbox{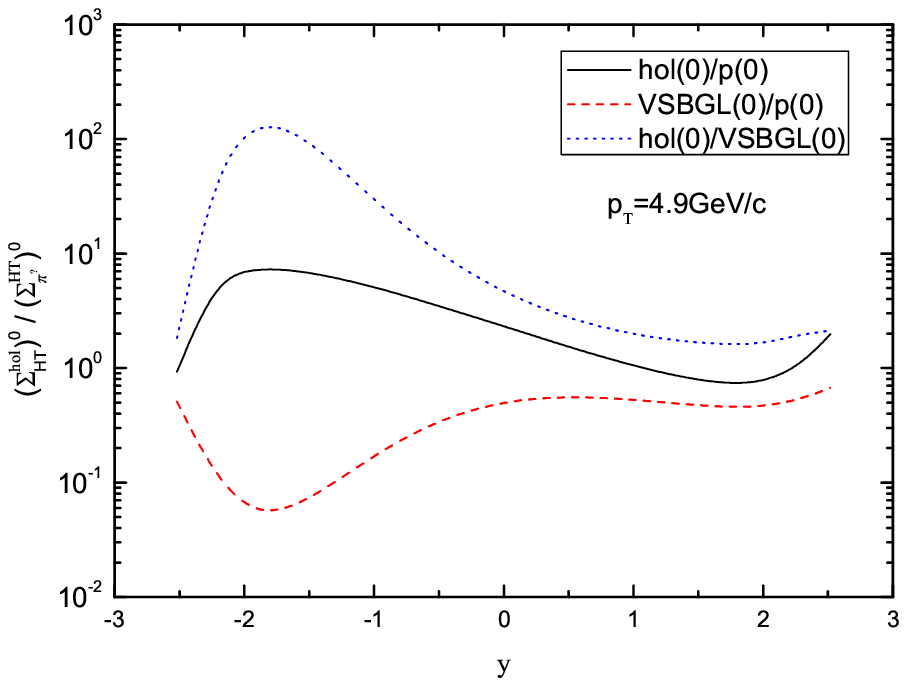}}
\vskip-0.2cm \caption{Ratio
$(\Sigma_{HT}^{hol})^{0}/(\Sigma_{\pi^{+}}^{HT})^0$, as a function
of the $y$ rapidity of the pion at the  transverse momentum of the
pion $p_T=4.9\,\, GeV/c$, at the c.m. energy $\sqrt s=62.4\,\,
GeV$.} \label{Fig9}
\end{figure}

\begin{figure}[!hbt]
\vskip-1.2cm \epsfxsize 11.8cm \centerline{\epsfbox{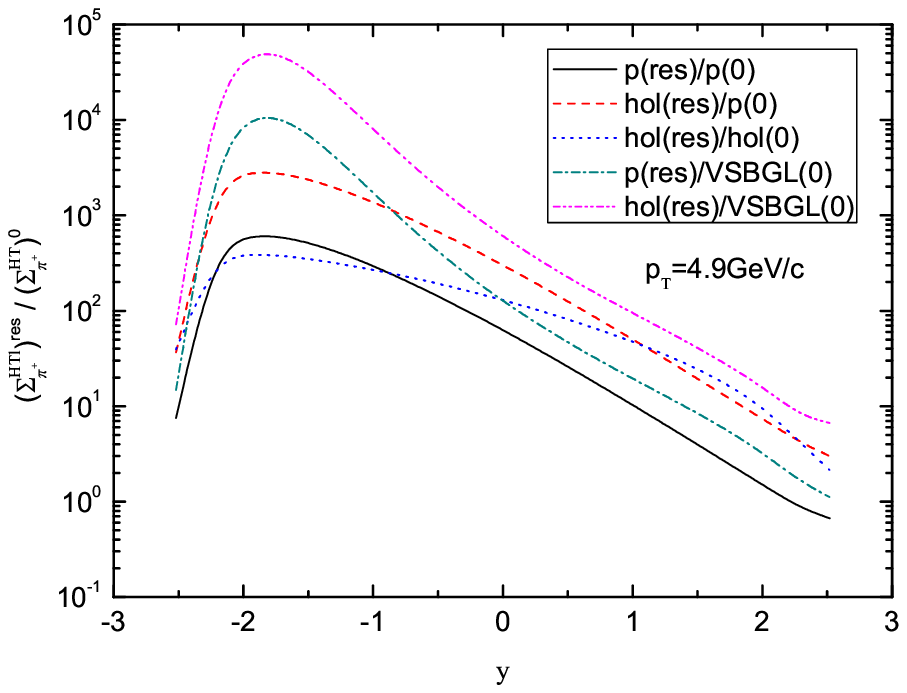}}
\vskip-0.2cm \caption{Ratio
$(\Sigma_{\pi^{+}}^{HT})^{res}/(\Sigma_{\pi^{+}}^{HT})^{0}$, as a
function of the $y$ rapidity of the pion at the  transverse momentum
of the pion $p_T=4.9\,\, GeV/c$, at the c.m. energy $\sqrt
s=62.4\,\, GeV$.} \label{Fig10}
\end{figure}

\begin{figure}[!hbt]
\vskip-1.2cm\epsfxsize 11.8cm \centerline{\epsfbox{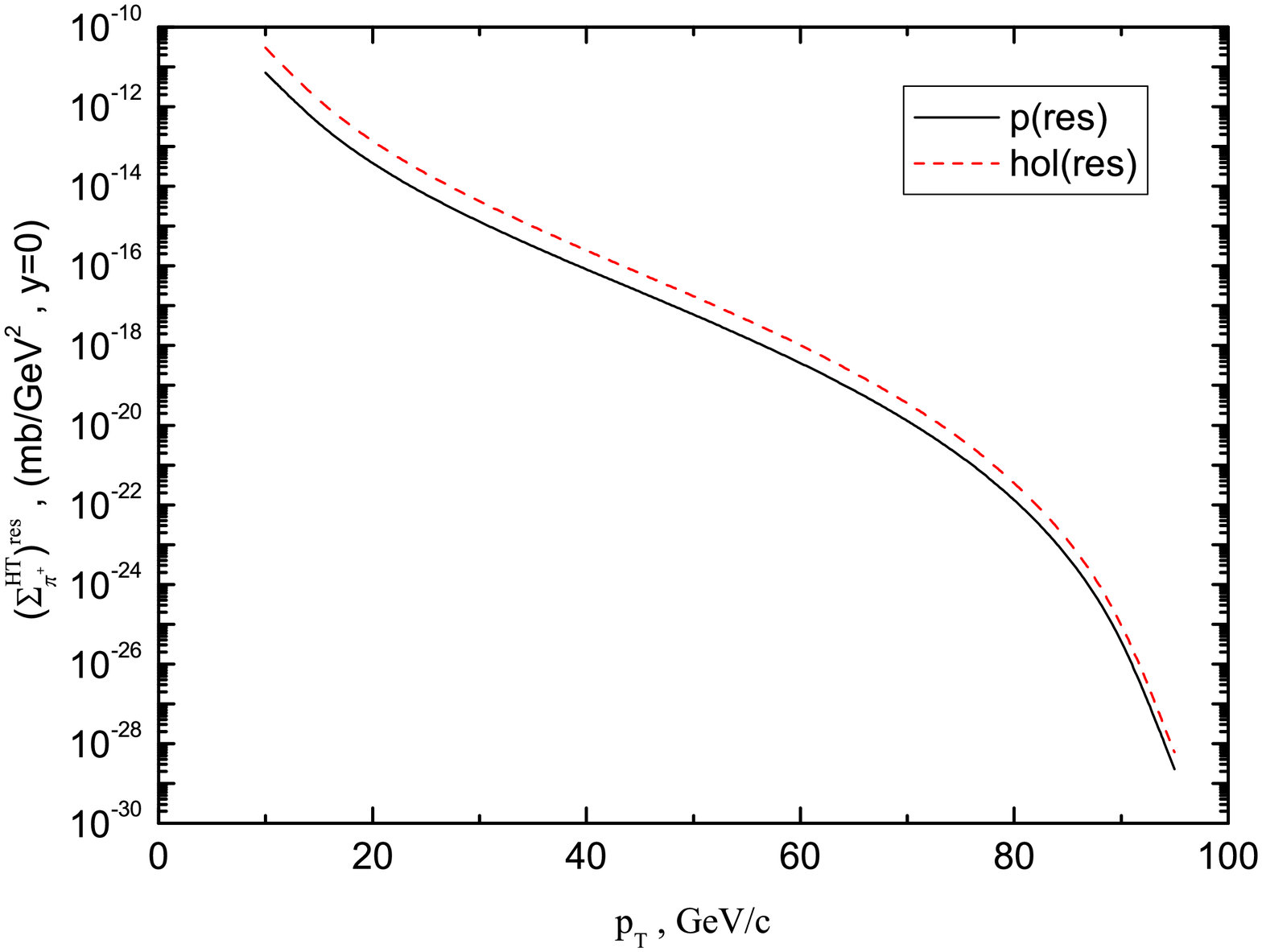}}
\vskip-0.2cm \caption{Higher-twist $\pi^{+}$ production cross
section $(\Sigma_{\pi^{+}}^{HT})^{res}$ as a function of the $p_{T}$
transverse momentum of the pion at the c.m.energy $\sqrt s=200\,\,
GeV$.} \label{Fig11}
\end{figure}

\begin{figure}[!hbt]
\epsfxsize 11.8cm \centerline{\epsfbox{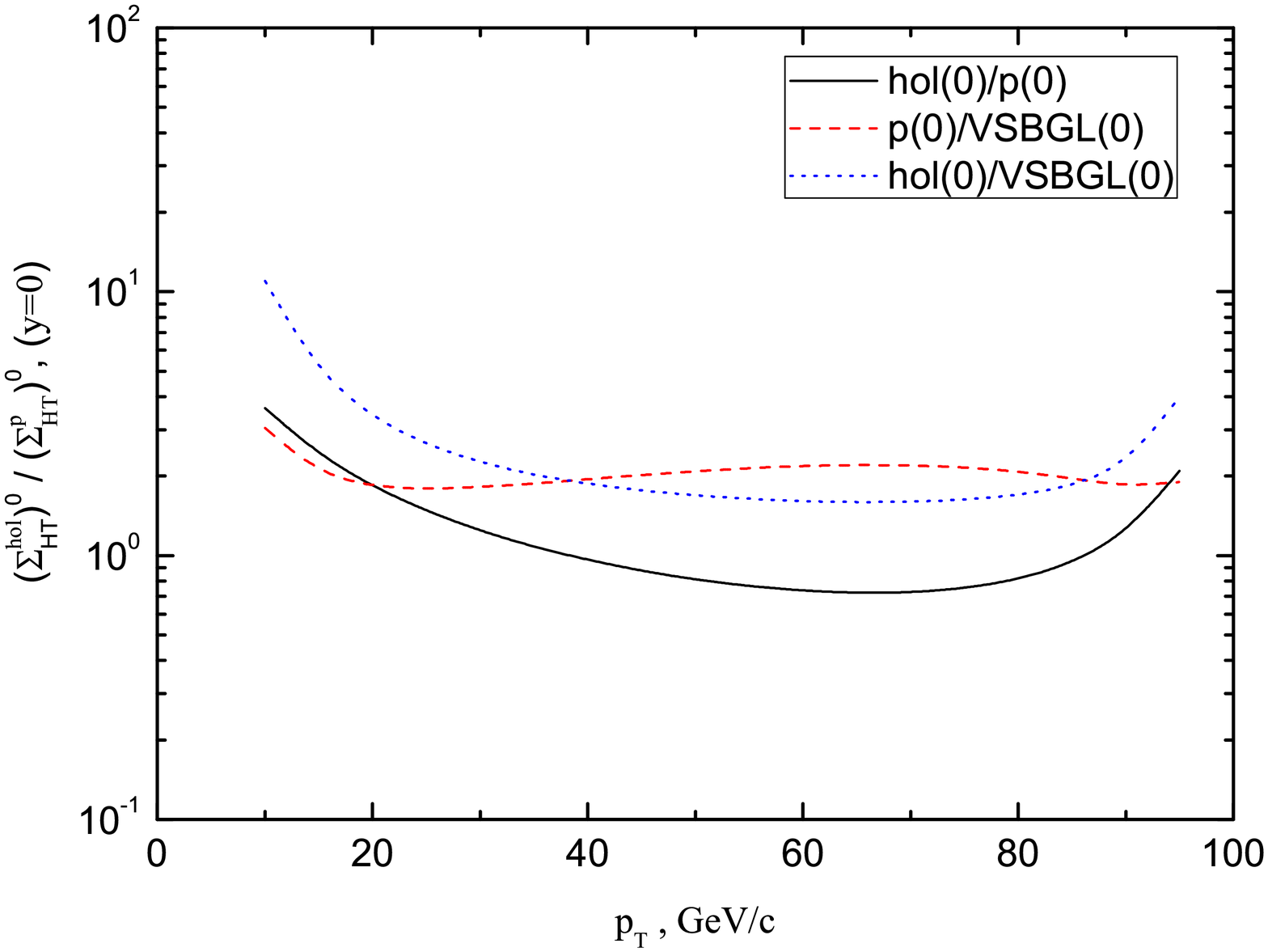}} \vskip-0.05cm
\caption{Ratio $(\Sigma_{HT}^{hol})^{0}/(\Sigma_{HT}^p)^{0}$, as a
function of the $p_{T}$ transverse momentum of the pion at the
c.m.energy $\sqrt s=200\,\, GeV$.} \label{Fig12}
\end{figure}

\begin{figure}[!hbt]
\vskip-1.2cm\epsfxsize 11.8cm \centerline{\epsfbox{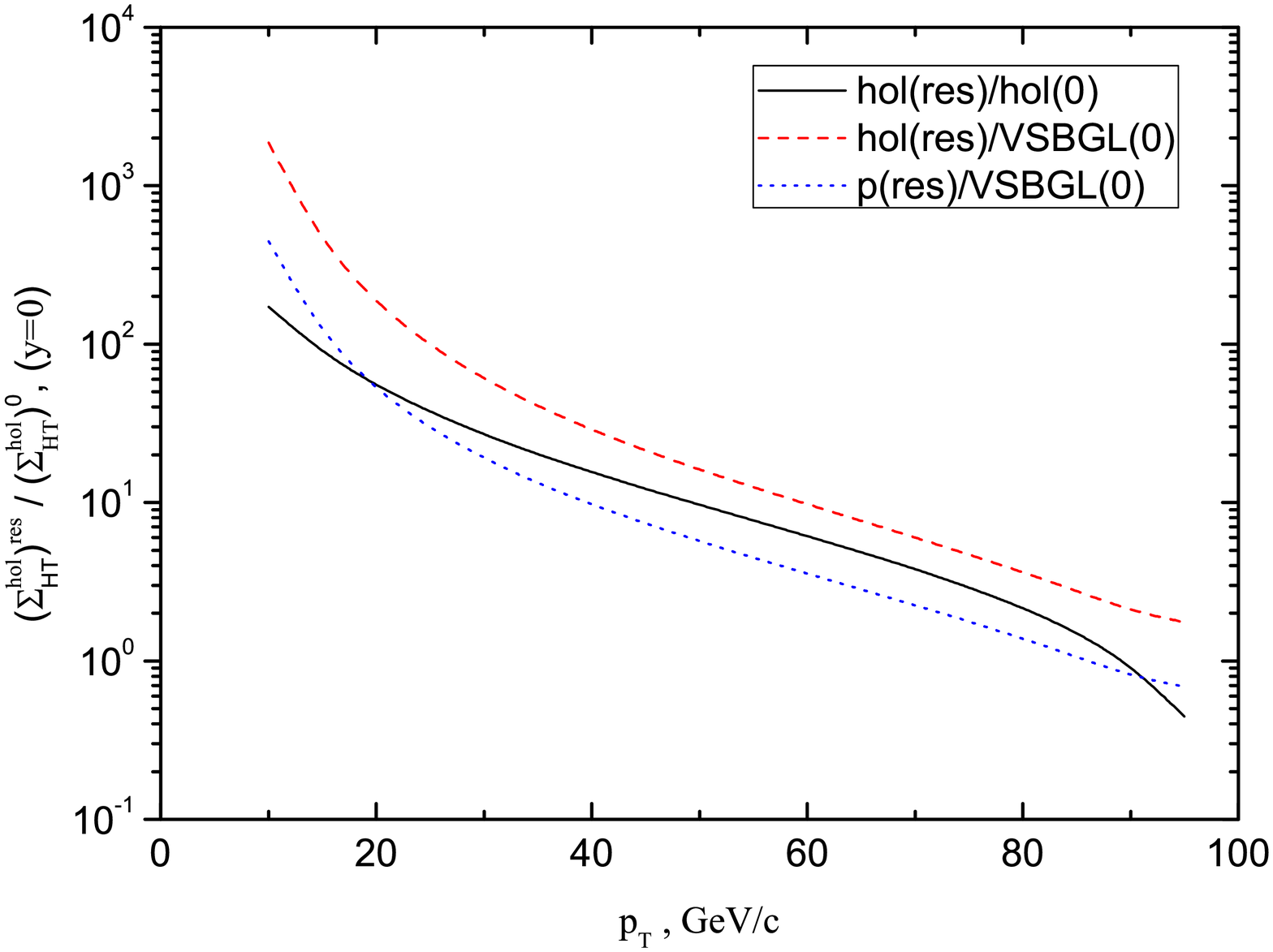}}
\vskip-0.2cm \caption{Ratio
$(\Sigma_{HT}^{hol})^{res}/(\Sigma_{HT}^{hol})^{0}$, as a function
of the $p_{T}$ transverse momentum of the pion at the c.m.energy
$\sqrt s=200\,\,GeV$.} \label{Fig13} \vskip 1.8cm
\end{figure}

\clearpage
\begin{figure}[!hbt]
\vskip 0.8cm \epsfxsize 11.8cm \centerline{\epsfbox{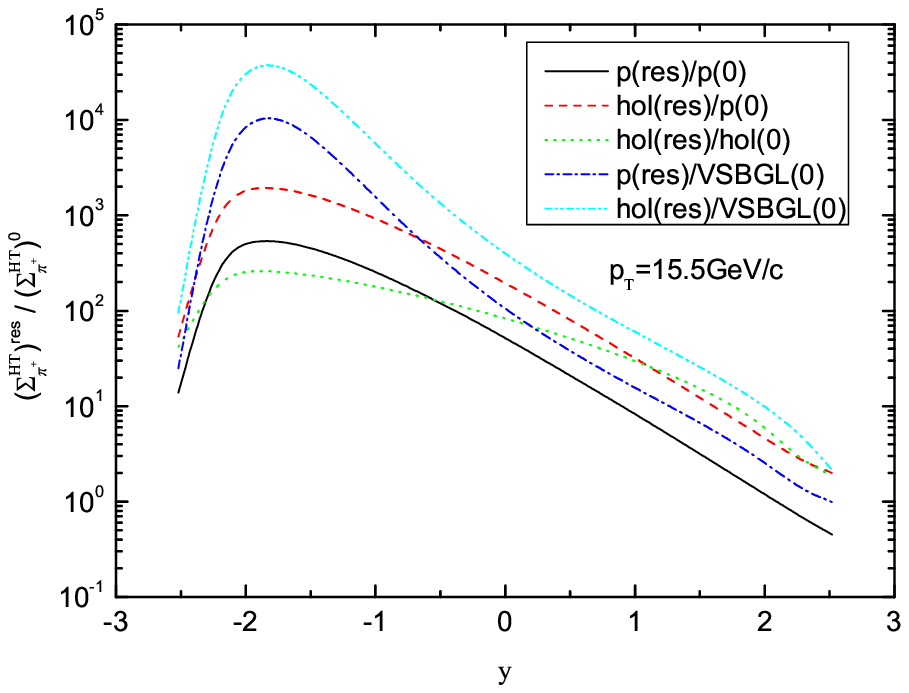}}
\vskip-0.2cm \caption{Ratio
$(\Sigma_{\pi^{+}}^{HT})^{res}/(\Sigma_{\pi^{+}}^{HT})^{0}$, as a
function of the $y$ rapidity of the pion at the  transverse momentum
of the pion $p_T=15.5\,\, GeV/c$, at the c.m. energy $\sqrt
s=200\,\, GeV$.} \label{Fig14}
\end{figure}

\end{document}